\documentclass[final,3p,times]{elsarticle}
\usepackage{lineno,hyperref}
\modulolinenumbers[5]
\usepackage{xcolor}

\usepackage{soul}
\usepackage{rotating}
\usepackage{booktabs}
\journal{Chaos, Solitons \& Fractals}

\biboptions{numbers,sort&compress}

\begin{document}

\begin{frontmatter}

\title{Forecasting countries' gross domestic product from patent data}

\author[address1,address2]{Yucheng Ye \fnref{myfootnote}}
\fntext[myfootnote]{These authors contributed equally to this work.}
\author[address1,address2]{Shuqi Xu\fnref{myfootnote}\corref{mycorrespondingauthor}}
\cortext[mycorrespondingauthor]{Corresponding author}
\ead{xushuqi@std.uestc.edu.cn}
\author[address1,address2,address3]{Manuel Sebastian Mariani}
\author[address1,address2,address4]{Linyuan L\"u\corref{correspondingauthor}}
\cortext[correspondingauthor]{Corresponding author}
\ead{linyuan.lv@uestc.edu.cn}

\address[address1]{Yangtze Delta Region Institute (Huzhou), University of Electronic Science and Technology of China, Huzhou 313001, PR China}
\address[address2]{Institute of Fundamental and Frontier Sciences, University of Electronic Science and Technology of China, Chengdu 611731, PR China}
\address[address3]{URPP Social Networks, Universit\"at Z\"urich, 8050 Z\"urich, Switzerland}
\address[address4]{Beijing Computational Science Research Center, Beijing 100193, PR China}

\begin{abstract}
Recent strides in economic complexity have shown that the future economic development of nations can be predicted with a single ``economic fitness'' variable, which captures countries' competitiveness in international trade. The predictions by this low-dimensional approach could match or even outperform predictions based on much more sophisticated methods, such as those by the International Monetary Fund (IMF). However, all prior works in economic complexity aimed to quantify countries' fitness from World Trade export data, without considering the possibility to infer countries' potential for growth from alternative sources of data.
Here, motivated by the long-standing relationship between technological development and economic growth, we aim to forecast countries' growth from patent data. Specifically, we construct a citation network between countries from the European Patent Office (EPO) dataset. Initial results suggest that the H-index centrality in this network is a potential candidate to gauge national economic performance. To validate this conjecture, we construct a two-dimensional plane defined by the H-index and GDP per capita, and use a forecasting method based on dynamical systems to test the predicting accuracy of the H-index. We find that the predictions based on the H-index-GDP plane outperform the predictions by IMF by approximately 35\%, and they marginally outperform those by the economic fitness extracted from trade data. Our results could inspire further attempts to identify predictors of national growth from different sources of data related to scientific and technological innovation.
\end{abstract}


\end{frontmatter}


\section{Introduction}

Forecasting the gross domestic product (GDP) of countries is a central problem in economic forecasting. Extant approaches fall into two categories. Forecasts such as those by the International Monetary Fund (IMF) take into account a large number of potential predictors, and rely on country-specific assumptions made by country-specific experts\footnote{\url{https://www.imf.org/external/pubs/ft/weo/faq.htm##q1g}}. By contrast, the recent \textit{economic complexity} approach has shown that accurate forecasts can be achieved by capturing the competitiveness of a country with a single variable, referred to as \textit{economic fitness}~\cite{tacchella2012new}. This approach has produced a stream of works that aimed to infer countries' economic fitness from World Trade export data, under the fundamental assumption that a competitive country is the one that exports many products, including those exported by few other countries~\cite{tacchella2012new,mariani2015measuring,wu2016mathematics,sciarra2020reconciling} -- see~\cite{mariani2019nestedness} for a recent review. Importantly, GDP predictions based on the economic fitness variable produce forecasts whose accuracy is comparable with or superior to that of the more complex predictions by the IMF~\cite{cristelli2017predictability,tacchella2018dynamical}. Despite these encouraging findings, in the economic complexity field, there have not been studies attempting to quantify countries' economic fitness from sources of data other than World Trade export data. 

Here, we aim to forecast countries' GDP from technological centrality metrics derived from patent data. GDP predictions based on patent data can be beneficial for two main reasons. From a theoretical perspective, they would show that the relationship between countries' economic and technological development is not only a static correlation (see literature review below), but it holds predictive value. From an economic complexity perspective, these predictions would demonstrate that countries' potential for growth could be also inferred with simple indicators that do not require export data. Compared to prior findings based on export data~\cite{tacchella2018dynamical,fan2021rise}, accurate predictions based on diverse data sources may potentially lead to a broader set of predictive indicators, which may increase the accuracy and robustness of the resulting forecasts.

To uncover the predictive value of patent data, we use a patent dataset from the European Patent Office (EPO) over 35 years (1978--2015) to create a country-country citation network. Departing from prior works that measured countries' technological competitiveness via their number of patents~\cite{archibugi1992patenting, ernst2003patent, frietsch2006technological, nesta2004national, su2011building}, we measure the countries' centrality in the resulting citation network via various network-based metrics. We find that the network centralities have a significantly positive correlation with both the GDP and GDP per capita, and the highest correlation with the GDP per capita is achieved by the H-index. The observed correlation is relatively stable over time. Motivated by this result, inspired by a dynamical-systems approach to GDP forecasting~\cite{cristelli2015heterogeneous,cristelli2017predictability,tacchella2018dynamical}, we construct a two-dimensional dynamical plane, defined by the H-index and the GDP per capita (HI-GDP). In the plane, we observe that countries tend to improve their H-index over time and, more importantly, the plane exhibits regions where the dynamics are highly predictable. Motivated by this finding, we test the predictive power of the Selective Predictability Scheme (SPS)~\cite{tacchella2018dynamical} method based on the HI-GDP plane.
The resulting predictions outperform those by the IMF by more than 35\% (in terms of reducing the error rate), and they marginally outperform those by the economic fitness approach. Our findings not only provide empirical evidence for the predictability of national economic development from the strength of national technological development, but they also contribute to the economic complexity literature on macroeconomic predictions from the perspective of the network and complexity science. 

\section{Literature review}

\subsection{Technological development and economic growth}

The correlation between technological development and economic production has been studied for a long time in economics~\cite{cobb1928theory,solow1956contribution,romer1986increasing,lucas1988mechanics,romer1990endogenous, dosi2002technology,rosenberg2020technology}.
Among them, in 1927, 
Cobb et al.~\cite{cobb1928theory} introduced the technique into the production function to
describe the relationship between a country's gross industrial output and the combination of technological capability, labor and capital inputs.
In 1957, Solow et al.~\cite{solow1956contribution} introduced a neoclassical economic growth model that emphasized technological progress. Their model shows that economic growth not only depends on the growth of capital and labor, but also on technological progress. 
In the late 1980s, the ``New Growth Theory'' emerged~\cite{romer1986increasing,lucas1988mechanics}, 
theorizing that economic growth is the product of endogenous technological change. In other words, the fundamental reason for sustained economic growth is that knowledge spillovers and human capital investment lead to technological change and iterative renewal of the whole society.
In 1990, Romer et al.~\cite{romer1990endogenous} proposed an endogenous growth model of technological progress, theorizing that economic growth builds on scientific and technological progress. 
Although there are considerable efforts on the correlation between technological development and economic production, we lack studies demonstrating the predictive power of the observed correlations.

Thanks to the increasing availability of high-resolution data on countries' economic activities, researchers have the opportunity to move from the above theories to the validation in practice,
which provided empirical evidence for the positive relationship between national technological development and economic growth~\cite{adak2015technological,pece2015innovation,shanmuganathan2008modelling,wong2005entrepreneurship,bernardes2006modeling,jorgenson2001information,dedrick2003information,ulku2004r}.
For example, 
Adak et al.~\cite{adak2015technological} found a positive correlation between the number of patent applications in Turkey and the country's GDP growth.
Pece et al.~\cite{pece2015innovation} used metrics such as the number of patents, the number of trademarks, and R\&D expenditures to quantify countries' technological competitiveness. 
They used multiple regression models to analyze the data of three Central-Eastern European countries, and found a positive connection between economic growth and technological competitiveness.
The above two studies and related works lack a predictive analysis that shows whether variables related to technological development can forecast the future GDP of countries.
In this respect, the work of Bernardes et al.~\cite{bernardes2006modeling} went one step further.
The authors collected paper and patent data for 183 countries from 1999 to 2003, as well as the population and GDP data in 2003, and built a model to simulate a world economy which has a high correlation with the real world, revealing the positive correlations between scientific activities, technological innovation, and national income.

However, these studies typically focus on complex models with many statistical indicators based on origin data like patents, and they do not describe the joint dynamics of technological and economic development.
Different from the existing literature, our paper aims to forecast countries' future GDP from network-based indicators of technological impact.

\subsection{Economic complexity}

Recently, a series of papers introduced country-level competitiveness variables based on the country's position in the country-product bipartite network inferred from World Trade export data~\cite{hidalgo2009building,tacchella2012new,mariani2015measuring,wu2016mathematics,sciarra2020reconciling} -- see~\cite{mariani2019nestedness} for a review. The core idea of this approach is that a country's intangible capabilities determine the set of products the country can produce and export; as a result, we can infer a country's capabilities by analyzing the diversification of its exported products~\cite{cristelli2013measuring}.
Among them, Tacchella et al.~\cite{tacchella2012new} developed a non-monetary metric named \textit{fitness} based on a non-linear iterative algorithm. 
Motivated by the goal of economic forecasting, Cristelli et al.~\cite{cristelli2015heterogeneous} uncovered the laminar-like regions of countries' dynamics in the fitness-GDP plane; narrower laminar regions were observed for alternative network-based metrics~\cite{liao2017ranking}. Based on the findings, the authors proposed a method called Selective Predictability Scheme from the perspective of dynamical systems (see Sec.~3.4 for details) to predict countries' future evolution in the plane. Besides, Tacchella et al.~\cite{tacchella2018dynamical} proved that the result of this approach is more than 25\% more accurate than the five-year forecasts published by the IMF. 
Because of its predictive value, the economic fitness has been incorporated among the World Bank indicators\footnote{\url{https://datacatalog.worldbank.org/dataset/economic-fitness}}, and the economic complexity approach has been further extended to analyze a broader range of problems, including the green economy~\cite{sbardella2018green}, economic inequalities~\cite{sbardella2017economic}, and the importance of services for countries' competitiveness~\cite{zaccaria2018integrating}, among others.
To date, studies on GDP prediction based on an economic complexity approach focused on inferring countries' fitness from export data and using the resulting variable to make GDP predictions. Departing from the focus on export data, we aim to identify predictors of countries' future GDP from patent data.

\section{Methods}

\subsection{Patent data and economic indicators}

We analyze the patent dataset from the European Patent Office (EPO) over the period 1978 - 2015, which consists of 1,378,320 patents and 885,721 citations. 
There are three ways to attribute a patent to a country~\cite{van2001using}: (1) by the country of the location of the applicant; (2) by the country of the residence of the inventor; (3) by the country of application priority\footnote{When a patent is filed multiple times in multiple countries, the country of application priority is the first one in which the patent was originally filed. }.
Considering the applicant (usually a firm, sometimes a government body or individual) is the patentee of the application, we choose the first approach to define the country that a patent belongs to~\cite{van2001using}. In the analyzed EPO data, the applicants of patents come from 131 countries, involving 43 European countries/regions, 82 non-European countries/regions, and 6 dissolved states or patent organizations (see for Table~A1 for a full list of them). 

One caveat for using this kind of local patent office database is that not all countries' patenting activities are adequately represented in the data, given that domestic applicants tend to apply for more patents in their home country compared to foreign applicants. As one would expect, European applicants prefer to apply for patents in European Patent Office, whereas US applicants prefer the US Patent Office, and Chinese applicants tend to be most represented in the Chinese Patent Office. This is called the \textit{home bias effect} in the literature~\cite{criscuolo2006home}. For the EPO data analyzed here, the patenting records of non-European countries might underrepresent their technological innovation activities. To mitigate the potential home bias, in the correlation analysis below, we consider separately European and non-European countries. However, to compare the resulting GDP predictions against state-of-the-art predictions, we include all the countries in the analysis -- with the filters applied and explained below, we obtain growth forecasts from the EPO dataset for about 70 countries and regions. 

To assess a country's level of economic development, there are many available macroeconomic indicators. The main indices include Gross Domestic Product (GDP), Gross National Income (GNI), Production Price index (PPI), Consumer price index (CPI), and so on. In line with previous works in the economic complexity literature~\cite{cristelli2015heterogeneous,cristelli2017predictability,tacchella2018dynamical}, we consider the problem of forecasting countries' GDP per capita to analyze how the economy is growing with its population. This choice is also based on the fact that technology is an essential factor that helps countries to improve their per capita performance with a stable population level~\cite{ogburn1959technological,teitel1994patents}. Specifically, we use the GDP pc based on purchasing power parity in constant 2017 international dollars (GDP, PPP (constant 2017 international \$)) in the analysis, which is available at \url{https://data.worldbank.org/}.
Motivated by previous findings~\cite{cristelli2015heterogeneous,tacchella2018dynamical}, we also consider a non-monetary metric, the economic \textit{fitness}~\cite{tacchella2012new}, as the state-of-the-art method to infer the growth potential of countries based on worldwide trade data. The economic fitness dataset is available at \url{https://datacatalog.worldbank.org/dataset/economic-fitness}.

\subsection{Country-country citation network}

To measure the countries' technological impact, we build a citation network between countries based on the patent dataset mentioned above. Specifically, if patent $i$ cites patent $j$, and the applicant of patent $i$ resides in country $A$, the applicant of patent $j$ comes from country $B$, then the link $A\rightarrow B$ is created. Self-citations from a given country to the same country are excluded. It is possible that a patent relates to several applicants from different countries. To fairly treat multiple links associated with a single patent citation, we give each link the weight equal to $1/(mn)$, where $m$ and $n$ denote the number of countries that the applicants of the citing patent and cited patent reside in, respectively~\cite{radicchi2009diffusion,west2013author}. After building the initial weighted links between countries, we adopt a criterion---Revealed Comparative Advantage (RCA)~\cite{balassa1965trade}---to quantify countries' relative advantage reflected in patent citation relations, thus transform the weighted network into an unweighted one\footnote{Tacchella et al.~\cite{tacchella2012new} found that the correlation between fitness based on the country-product matrix and GDP pc can be enhanced when the matrix is processed according to the RCA rule. A similar result is also found in our study, in which after RCA processing, the correlation between the metrics from the unweighted network and GDP pc is higher than that from weighted network.}.
In detail, the RCA value is defined as the ratio between the fraction of observed citation number of two countries with respect to all observed citation number of the citing country and the fraction of all observed citation number of the cited country with respect to all the observed citation number. In the formula, the RCA for country $i$ citing country $j$ is:
\begin{equation}
\centering
RCA_{ij} = \frac{\frac{w_{ij}}{\sum_j w_{ij}}}{\frac{\sum_i w_{ij}}{\sum_{ij} w_{ij}}},
\end{equation}
where $w_{ij}$ is the link weight from country $i$ to country $j$ in the initial weighted network.
We consider only edges with $RCA \ge 1$, in this way, the original country-country citation network is transformed into a unweighted network represented by an adjacency matrix $A$ in which $A_{ij}=1$ if $RCA_{ij}\ge 1$ and 0 otherwise.

\begin{figure}[t]
\centering
\includegraphics[scale=0.35]{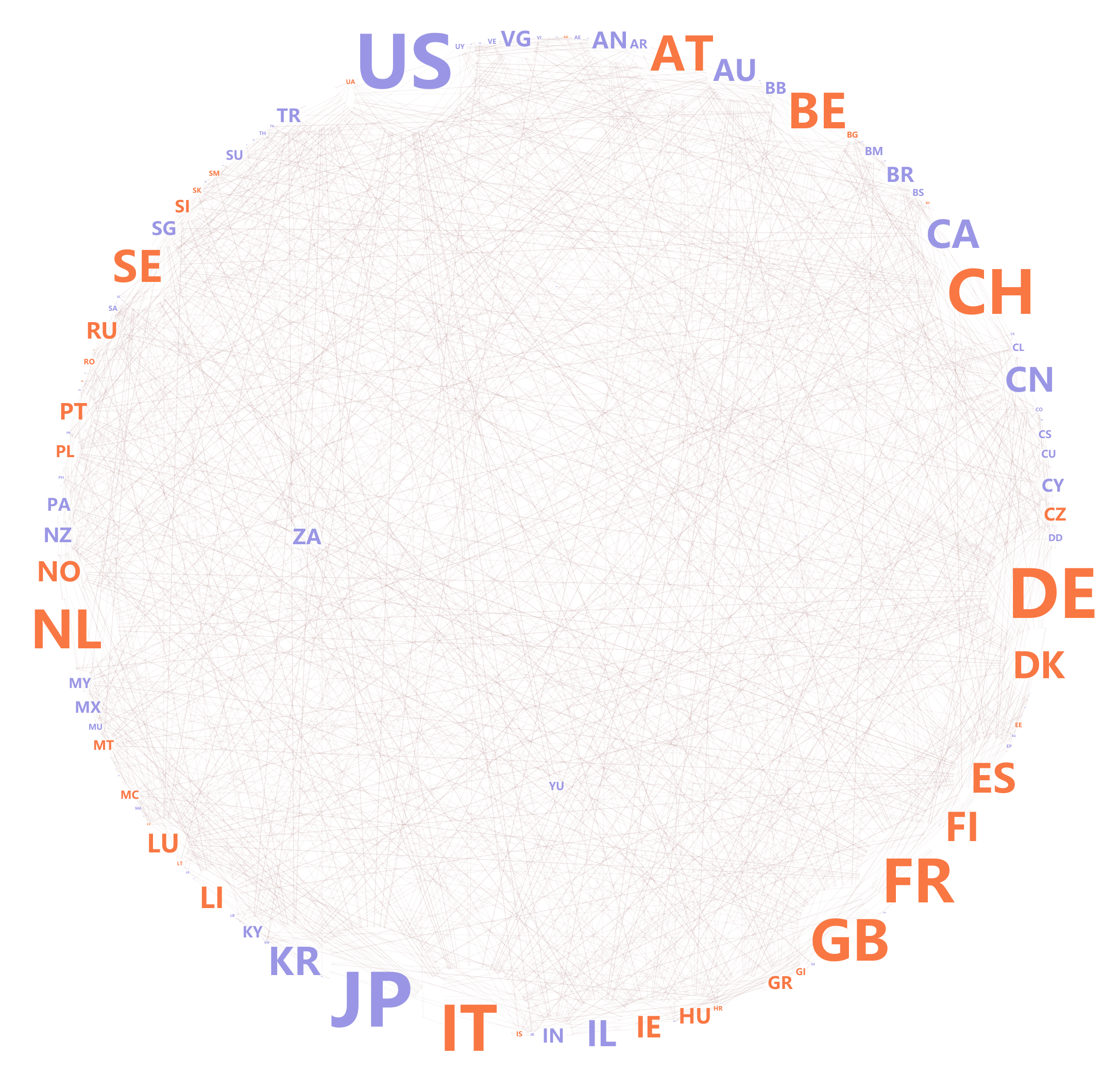} 
\caption{A visualization of the country-country citation network. The size of a country node is proportional to its indegree, i.e., how many countries have cited patents from this country. European and non-European countries are colored in orange and purple, respectively (see Table~A1 for all involved countries and their codes).}
\label{figure1}
\end{figure}

\begin{table}[ht]
    \centering
    \caption{The characteristics of the constructed unweighted citation network.}
    \begin{tabular}{ccccc}
        \toprule  
         Nodes & Edges & Average indegree& Density  \\
         \midrule
         131 & 1,284 & 9.802 &0.151\\
         \midrule
         Clustering coefficient & Indegree assortativity & Diameter & Average path length  \\
         \midrule
           0.391 & -0.455 & 4 & 2.102 \\
         \bottomrule
    \end{tabular}
    \label{tab:network}
\end{table}

The unweighted network includes 131 nodes (countries/regions) and 1,284 links, see Fig.~\ref{figure1} for its visualization. 
This network describes the citation relationship between (the agents residing in) countries, in which countries with a larger in-degree are cited by more countries. Figure~\ref{figure1} illustrates that Germany ($DE$), Italy ($IT$), France ($FR$) are the top three among European countries in terms of indegree. As for non-European countries, Japan ($JP$), the United States ($US$), and South Korea ($KR$) occupy the first three positions.
Table \ref{tab:network} reports the characteristics of the network. 
The network has no isolated node.
A country is cited by 9.802 countries on average, and 51 countries at most (\emph{US}).
This network also has a high-clustering feature ($C=0.391$) and a short average path length ($l=2.102$), which implies a small-world organization.
Besides, the constructed network shows disassortativity of indegree, meaning that countries with few citations tend to cite patents from countries with many citations.

\subsection{Countries' technological centralities}
Based on the country-country citation network, we consider five network centralities, including four local metrics: indegree~\cite{newman2012networks}, collective influence~\cite{morone2015influence, bovet2019influence}, H-index~\cite{hirsch2005index}, incoreness~\cite{lu2016h}, and one glocal metric: PageRank~\cite{brin1998anatomy}, to measure countries' technological centrality. Below, we provide their definitions. In the following section, we will investigate the correlation of these metrics with the GDP pc.

\textbf{Indegree}.
Indegree is a key metric to measure the importance of nodes in directed networks~\cite{liao2017ranking}. Based on the citation network constructed above, we calculate the indegree of country $i$ as $\sum_j A_{ji}$.

\textbf{Collective Influence (CI)}.
Collective influence has been introduced to identify the minimal set of influencers in networks via optimal percolation~\cite{morone2015influence, bovet2019influence}.
For country $i$, the \emph{$CI$} score at level $l$ is defined as~\cite{bovet2019influence}:
\begin{equation}
CI_i^l = (k_i-1)\sum_{j:d_{ij}=l}{(k_j-1)},
\end{equation}
where $k_i$ is the degree of country $i$,
and $d_{ij}$ denotes the shortest distance between country $i$ and country $j$, we use $l=2$ in this study.
Note that we use indegree in the computation of $CI$ for that indegree can represent the impact of nodes in the citation network.

\textbf{H-index}.
The H-index was originally proposed to evaluate researchers' academic achievement based on their publications and citations~\cite{hirsch2005index}.
L\"u et al.~\cite{lu2016h} abstracted the computation of the H-index as an operator $H$ that takes a set of real numbers $(x_1, x_2, \dots, x_n)$  as input and returns a maximum integer, $y=H(x_1, x_2, \dots, x_n)>0$, such that at least $y$ elements in $(x_1, x_2, \dots, x_n)$ are no less than $y$.
Then they obtained a family of H-indices defined by the equation: $h_i^{(n)} = H(h_{j_1}^{(n-1)}, h_{j_2}^{(n-1)},\ldots,h_{j_{k_i}}^{(n-1)})$. They found that the degree, H-index and coreness correspond to the initial state, intermediate state and steady state of the resulting family of $H$ indices respectively, and they pointed out that, for many cases in real networks, the H-index is a better measure of a node's influence than degrees or coreness.
In our study, for country $i$, the calculation of directed H-index is as:
\begin{equation}
\centering
h_i = H(k_{j_1}, k_{j_2},\ldots,k_{j_{k_i}}),
\end{equation}
where $k_i$ is the indegree of country $i$, and $k_{j_1}, k_{j_2},\ldots,k_{j_{k_i}}$ is the indegree of the countries citing country $i$.

\textbf{Incoreness}.
Coreness is a widely-used metric in vital node identification studies~\cite{lu2016vital, lu2016h}. 
The calculation of coreness is based on the well-known k-shell decomposition~\cite{seidman1983network}, proposed by Seidman in 1983.
The algorithm could be implemented through the following process. 
Firstly, delete all nodes with degree one, and then the pruning is repeated until all nodes' degrees are larger than one, and the removed nodes are assigned to the 1-shell and their coreness is one.
Next, repeat the pruning process for the rest nodes with degree $k=2$ and we will get the 2-shell of the network, where the deleted nodes have the coreness of 2. 
Continue to repeat the process, and all nodes get the corresponding coreness value.
An efficient implementation algorithm of determining the k-shell decomposition of a given network is proposed by Batagelj and Zaversnik~\cite{batagelj2003m}, which makes the time complexity of computing the coreness of nodes decreases to $O(m)$ where $m$ is the number of edges.
Here we modify the $core\_num$ function from the networkx library\footnote{\url{https://networkx.org/documentation/latest/reference/algorithms/generated/networkx.algorithms.core.core\_number.html\#networkx.algorithms.core.core\_number}} to make it consider the indegree of nodes.

\textbf{PageRank}.
PageRank was originally introduced to rank web pages in the World Wide Web~\cite{brin1998anatomy}, it has then found a broad range of applications in nodes ranking in many real systems~\cite{ding2009pagerank,gleich2015pagerank,ivan2011web,radicchi2011best,liao2017ranking}. For country $i$, the PageRank score is defined through the iterative equation as~\cite{liao2017ranking}:
\begin{equation}
\centering
P_i^{(n+1)} = \alpha\sum_{j:k^{out}>0}{\frac{A_{ji}}{k_j^{out}}P_j^{(n)}} +\alpha\sum_{j:k^{out}=0}{\frac{P_j^{(n)}}{N}}+\frac{1-\alpha}{N},
\end{equation}
where $k_j^{out}$ is the outdegree of country $j$, $\alpha$ is a damping factor with a default value of 0.85~\cite{liao2017ranking}, and $n$ is the iteration number. 
Note that when satisfying $\sum_{i=1}^N|P_i^{(n)}-P_i^{(n-1)}|/N<\epsilon$ where $\epsilon=10^{-9}$, we stop the iterative process.

\subsection{Selective Predictability Scheme (SPS)\label{sec:sps}}

Selective Predictability Scheme is a prediction approach proposed by Cristelli et al.~\cite{cristelli2015heterogeneous} to forecast countries' future dynamics in a two-dimensional plane. The underlying idea of this scheme is the method of \textit{analogues}~\cite{lorenz1969atmospheric, lorenz1969three}, which is effective when for a given system in exam, multiple longitudinal observations are available, while the general evolution rule is unknown~\cite{cristelli2015heterogeneous}. Specifically, Cristelli et al. built a two-dimensional plane, with one dimension representing the economic fitness, and the other dimension representing national income represented by GDP pc. To obtain a coarse-grained demonstration of all evolution observations of countries, the authors construct a grid in the fitness-GDP plane with square boxes, and the moving trajectories of countries can be simplified as a set of vectors in the plane and located in specific boxes. To predict countries' future dynamics over a $\Delta t$ time horizon (e.g., $\Delta t=5$ years), the method estimates the empirical evolution distribution of each box by tracking all the countries that originated from the box and recording where they are located in after $\Delta t$ years. This step is called SPS training. All moving vectors originating from a given box will be combined to compute an average $\Delta$-displacement of this box by defining a displacement from the center of all vectors' starting positions to the center of their ending positions. After the SPS is trained, a country's future moves from time $t$ to $t+\Delta t$ in the plane can be predicted by summing the average $\Delta$-displacement of the box it located in at time $t$ to its position in the plane at $t$. The results of the SPS method are proved to be robust with the change in the size of boxes~\cite{cristelli2015heterogeneous}.
Here we apply the SPS method based on a plane constructed of the countries' technological metric and GDP pc.

\section{Results}
\subsection{Correlation between technological centralities and GDP per capita}

\begin{figure}[!ht]
\centering
\includegraphics[scale=0.3]{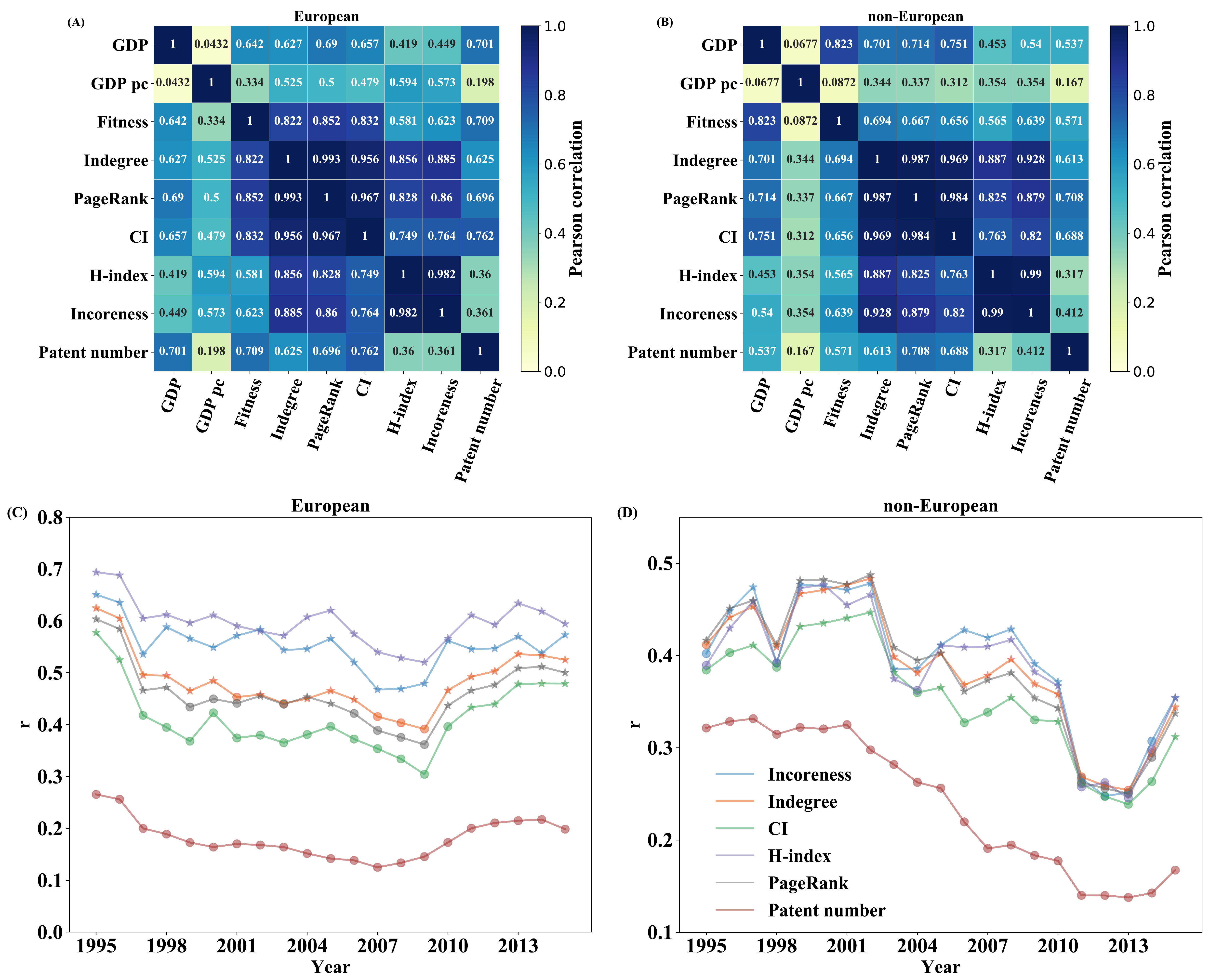} 
\caption{The correlation between patent metrics and GDP per capita. We study the European and the non-European countries independently. (A-B) The pairwise Pearson correlations among countries' GDP, GDP pc, fitness in 2015 and patent metrics based on the patent data until 2015. The results show that patent network metrics have high correlation with both GDP and GDP pc, and H-index has the highest correlation with GDP pc for European ($r=0.594$) and non-European ($r=0.354$) countries. (C-D) The Pearson correlation between patent metrics and GDP pc in different years. Significant ($\star$) refers to p-value less than 0.01. The correlation between H-index and GDP pc remains the highest among all metrics for European countries, while for non-European countries, H-index is close to the best metric. In addition, countries' patent number, which is a non-network metric, has the lowest correlation with GDP pc compared with network centrality metrics.}
\label{figure2}
\end{figure}

The pairwise Pearson correlation between all metrics computed in 2015 is shown in Fig.~\ref{figure2} (A-B). The results demonstrate that there is a weak correlation between the GDP and GDP pc, which means countries could perform differently in terms of the two measures. 
And the network centrality metrics have a positive correlation with both GDP and GDP pc, as well as fitness, which confirms that technological impact is positively associated with the economy. When calculating the patent metrics from the country citation network based on patents issued up to different years, the correlations between centrality metrics and GDP pc are relatively stable over time, see Fig.~\ref{figure2} (C-D). Comparing the network metrics with non-network metrics (i.e., patent number), we find that the network-based metrics exhibit a stronger correlation with the GDP pc than the non-network metrics for both European and non-European countries. This is because network centralities take into account not only how many patents does one country has, but also the citation relationship between countries and, therefore, technological impact~\cite{jaffe2019patent}. In general, for European countries, the H-index has a significantly higher correlation with the GDP pc (Pearson's $r=0.594$, see Fig.~\ref{figure2} (A))
compared to the other network-based metrics, and the correlation remains statistically significant over time, see Fig.~\ref{figure2} (C); on the other hand, for non-European countries, there is a negligible difference between the H-index and other centrality metrics, such as Incoreness and Indegree, see Fig.~\ref{figure2} (B,D).
Motivated by the stronger correlation with GDP pc exhibited by the H-index compared to other centralities, in the following, we mainly focus on countries' dynamics in terms of their H-index and GDP pc. 

\subsection{Dynamics in the plane of technological centrality and GDP per capita\label{sec:planedynamic}}

\begin{figure}[htbp]
\centering
\includegraphics[scale=0.35]{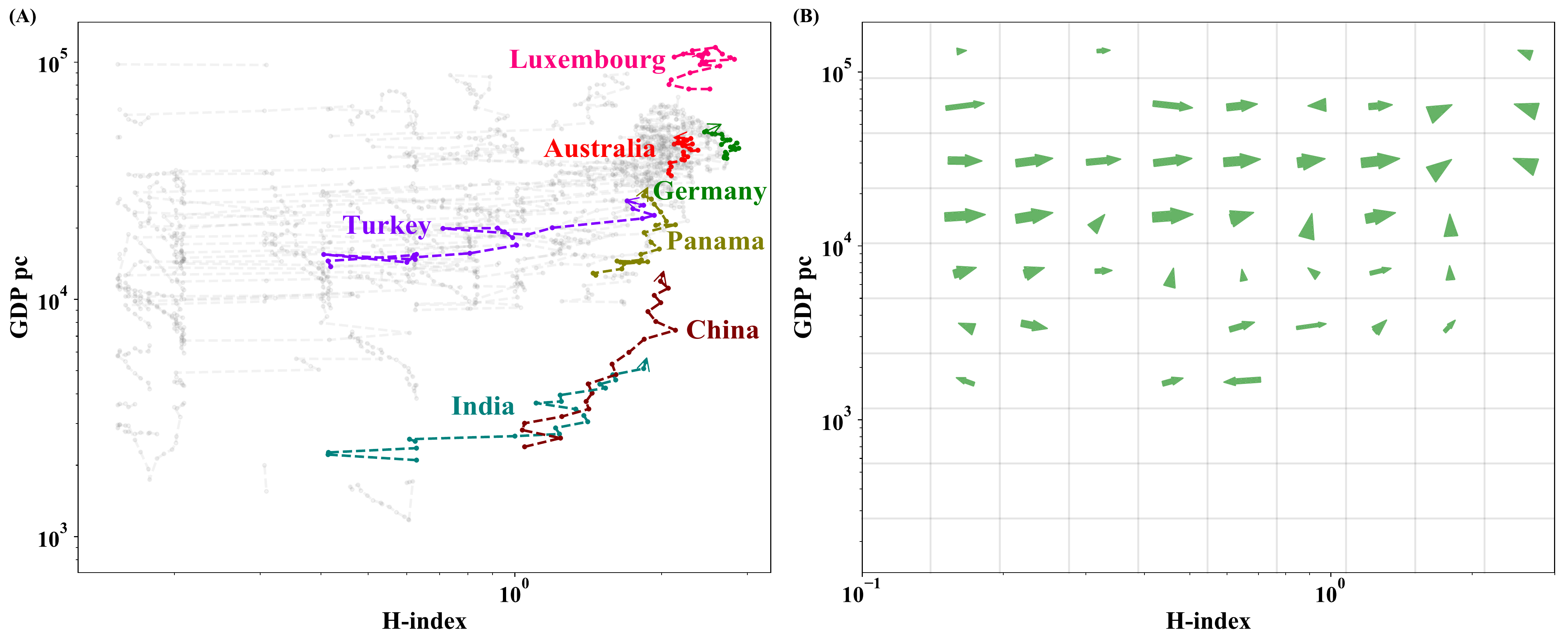}
\caption{Joint dynamics of countries' H-index and GDP per capita. 
(A) We show the evolution of countries in the HI-GDP plane from 1995 to 2015, several countries with clear upward trends are highlighted with colors. The specific H-index values of countries in each year are reported in Table \ref{tab:norm_H} in Appendix. (B) The coarse-grained illustration of countries' dynamics from 1995 to 2015. The plane is divided into $10 \times 10$ equally-sized boxes on a logarithmic scale. For each box, we consider all the 1-year moves of countries that start from this box to compute and plot the mean 1-year displacement for these moves. 
The result shows that the overall flow is laminar in the HI-GDP plane. 
Note that we normalize the H-index in both figures by the average value of every year, whereas GDP pc is not normalized with the unit of dollar (\$).
}
\label{figure3}
\end{figure}

Inspired by recent works~\cite{cristelli2015heterogeneous,liao2017ranking,cristelli2017predictability,tacchella2018dynamical}, to uncover the potential predictive value of countries' technological centrality metrics, we study the countries' dynamics in a two-dimensional plane where one dimension represents the technological centrality (represented by H-index) the other one represents economic development (represented by the GDP pc). Specifically, we build an H-index-GDP pc (HI-GDP) plane and show countries' dynamics from 1995 to 2015, see Fig.~\ref{figure3} (A). As mentioned above, the H-index of each country in a given year $t$ is computed based on the country citation network built based on patents issued up to year $t$. This ensures that when using the H-index at time $t$ to make predictions over a future time window $(t,t+\Delta t)$, no data from the future are used for the prediction. We find that during the past 20 years, in general, most countries show an increasing trend in H-index but a slow upward trend in GDP pc, while some countries have made great progress in both aspects.  
For example, China's technological centrality has improved rapidly in the past 20 years and gradually caught up with the developed countries, which could due to its continuously increasing investments in scientific and technological innovation. In parallel, China achieved rapid growth in terms of GDP pc. Similarly, India, which placed technological innovation as the core of national development, realized improvement of both H-index and GDP pc. See the specific H-index values of countries in each year in Table \ref{tab:norm_H} in Appendix.

In Fig.~\ref{figure3} (B), we illustrate the coarse-grained evolution of all countries in the plane in terms of 1-year displacements, from 1995 to 2015. In detail, we divide the plane into $10 \times 10$ equally-sized boxes on a logarithmic scale. Then the starting positions of the movement trajectories for each country within one year can be located in specific boxes. For each box, we average all the 1-year moves of countries (which we call ``events'' later) that start from this box to obtain the mean displacement of this box, which is displayed as an arrow in Fig.~\ref{figure3} (B). The thickness of arrows is determined by the number of events contained in the corresponding box, the direction and length of each arrow are the average results of all events involved in that box. Note that only boxes with at least three events are considered and shown, and the same threshold is applied in the following analysis.
From Fig.~\ref{figure3} (B), we can qualitatively observe several regions of the plane where the flow is approximately laminar, meaning that adjacent countries move toward the same direction.
This suggests that countries' past trajectories in the H-index and GDP pc plane can be used to predict their future situation, which we test in the following. 

\begin{figure}[!ht]
\centering
\includegraphics[scale=0.33]{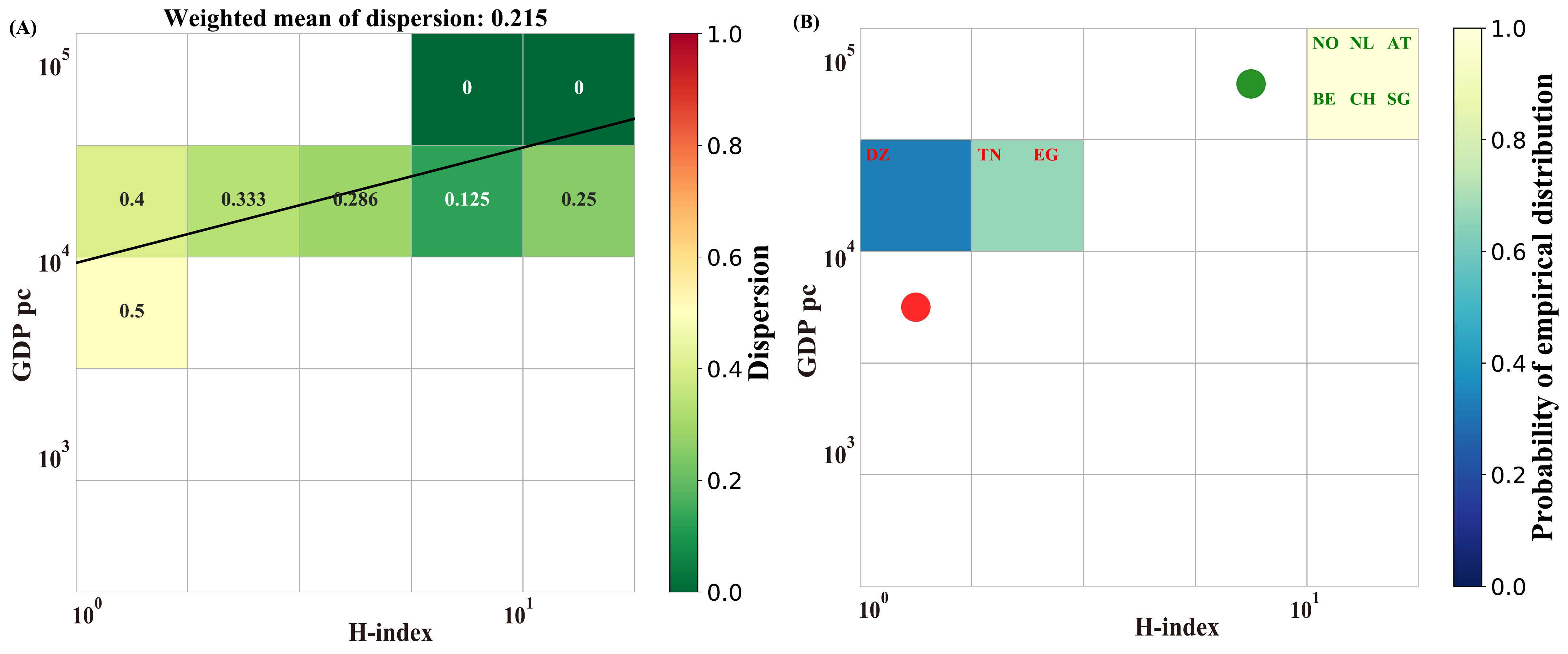}
\caption{Dispersion in the HI-GDP plane. (A) The box dispersion in the HI-GDP plane. The black line is obtained by fitting through all boxes with dispersion values by Ordinary Least Square method with $R^2=0.662$. The weighted mean of dispersion for all effective boxes with at least 3 events is 0.215. 
(B) The empirical distribution (ED) of the 10-year evolution for countries from two boxes: the box with a green dot has a low dispersion ($D=0$) while the box with a red dot has a relatively high dispersion ($D=0.5$). As for the former box, six countries from the box are concentrated in one box ten years later; while as for the latter box, the three countries located in this box transfer to two different boxes in the next 10 years.}
\label{figure4}
\end{figure}

To quantitatively analyze countries' dynamics in the HI-GDP plane, we refer to the number of countries located in box $i$ in year $t$ as $N_t^i$. For each box $i$, we track the evolution of countries within this box in the following 10 years and compute the number of different boxes that the tracked countries will occupy after 10 years (i.e., in year $t+10$), this number is labeled as $n_t^i$. 
In this way, the dispersion of box $i$ between year $t$ and year $t+10$ is computed as $(n_t^{i}-1)/(N_t^{i}-1)$~\cite{cristelli2015heterogeneous}. A box has zero dispersion if all countries within it will occupy the same box after 10 years (as shown in Fig.~\ref{figure4} (B)). Conversely, a box has dispersion equal to one if all the $N^i$ countries from it will occupy different $N^i$ boxes after 10 years. Therefore, smaller dispersion means lower uncertainty: the future evolution of countries located at the box with smaller dispersion is expected to be more predictable.

The dispersion result at $t=2005$ is shown in Fig.~\ref{figure4} (A). Further in Fig.~\ref{figure4} (B), we illustrate the evolution of countries from a box with a small dispersion ($D=0$) compared with countries from a box with a medium level of dispersion ($D=0.5$) over the following 10 years. Due to the limited number of countries in the dataset, in this part of the analysis, we consider both European and non-European countries.
In Fig.~\ref{figure4} (A), we observe that the dispersion of any box is no more than 0.5, and the differences are not significant, which suggest the possibility to predict countries' dynamics in the plane. The dispersion in the plane can be reduced to a weighted mean value, which weights each box according to the number of events in it~\cite{liao2017ranking}. The weighted mean of dispersion of the HI-GDP plane is 0.215. The findings are markedly different from the heterogeneous results found in the fitness and GDP pc plane~\cite{cristelli2015heterogeneous}, which reveal that the laminar regime only exists in the region with high fitness and low-fitness regions appear to be chaotic and exhibit a low level of predictability.

\begin{figure}[!ht]
\centering
\includegraphics[scale=0.33]{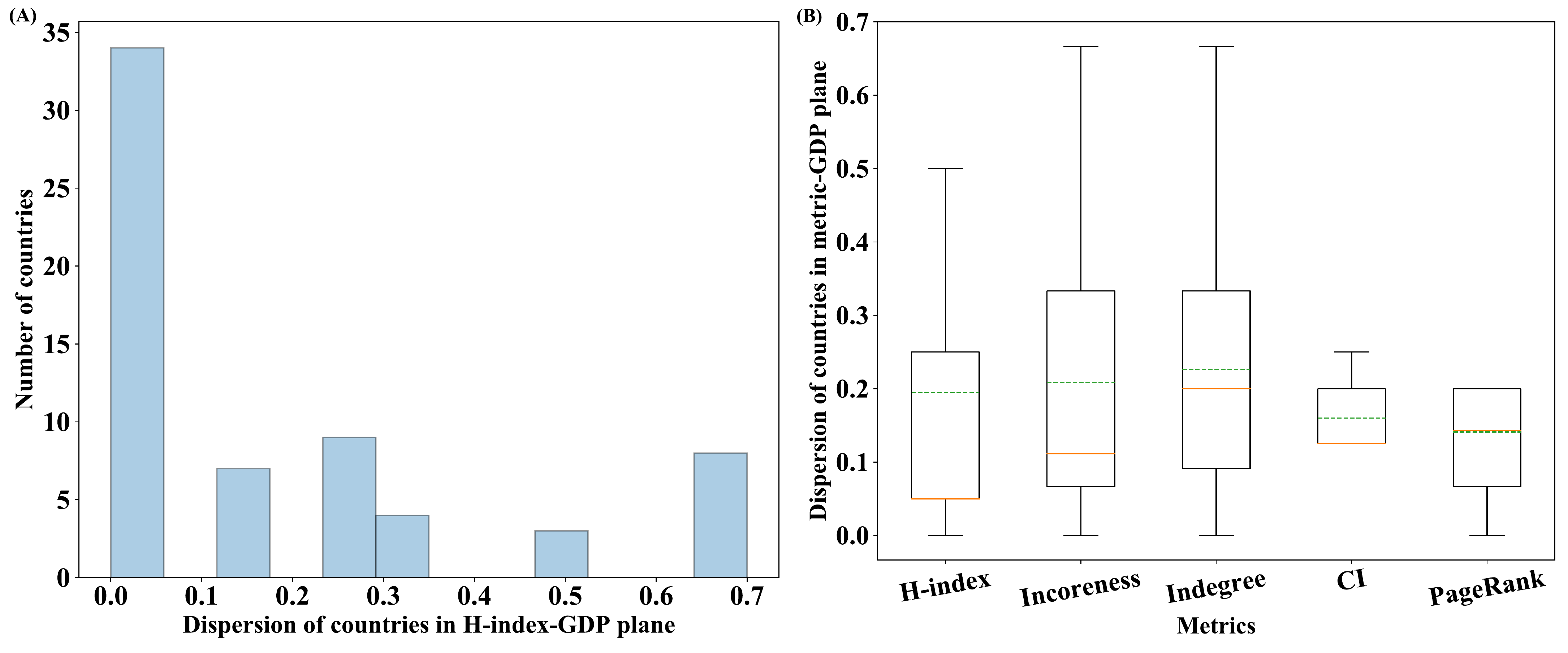} 
\caption{The distribution of the  10-year dispersion of countries in metric-GDP plane from 2005 to 2015.
(A) The distribution of countries' 10-year dispersion in the HI-GDP plane.
The left peak indicates that most countries exhibit low values of dispersion.
(B) The boxplot of the countries' 10-year dispersion in different metric-GDP planes, where the orange lines represent the median of dispersion, and the green dashed lines represent the mean of dispersion.}
\label{figure5}
\end{figure}

To analyze the predictability of countries in the HI-GDP plane, in Fig.~\ref{figure5} (A), we show the distribution of 10-year dispersion of countries in the HI-GDP plane from 2005 to 2015, where the dispersion of a given country over a specific period of time is defined as the dispersion of the box where the country situated at the start year. 
We find that, in the HI-GDP plane, more than half the countries exhibit low dispersion values, which suggests that the dynamics of most countries in the HI-GDP plane have high predictability. 
In Fig.~\ref{figure5} (B), we show the distribution of countries' dispersion values in different centrality metric-GDP planes, and we find that while the mean of dispersion values (the green dashed lines) for all planes is similar, the median of dispersion (the orange solid line) for the HI-GDP plane is significantly smaller than that for other planes, which denotes that the dynamics of countries in the HI-GDP plane are more predictable than those in other planes. This result also corresponds with the highest correlation of H-index with GDP pc among all patent metrics.

\subsection{Back-test analysis of the predictive power of the HI-GDP plane}

\begin{figure}[!ht]
\centering
\includegraphics[scale=0.33]{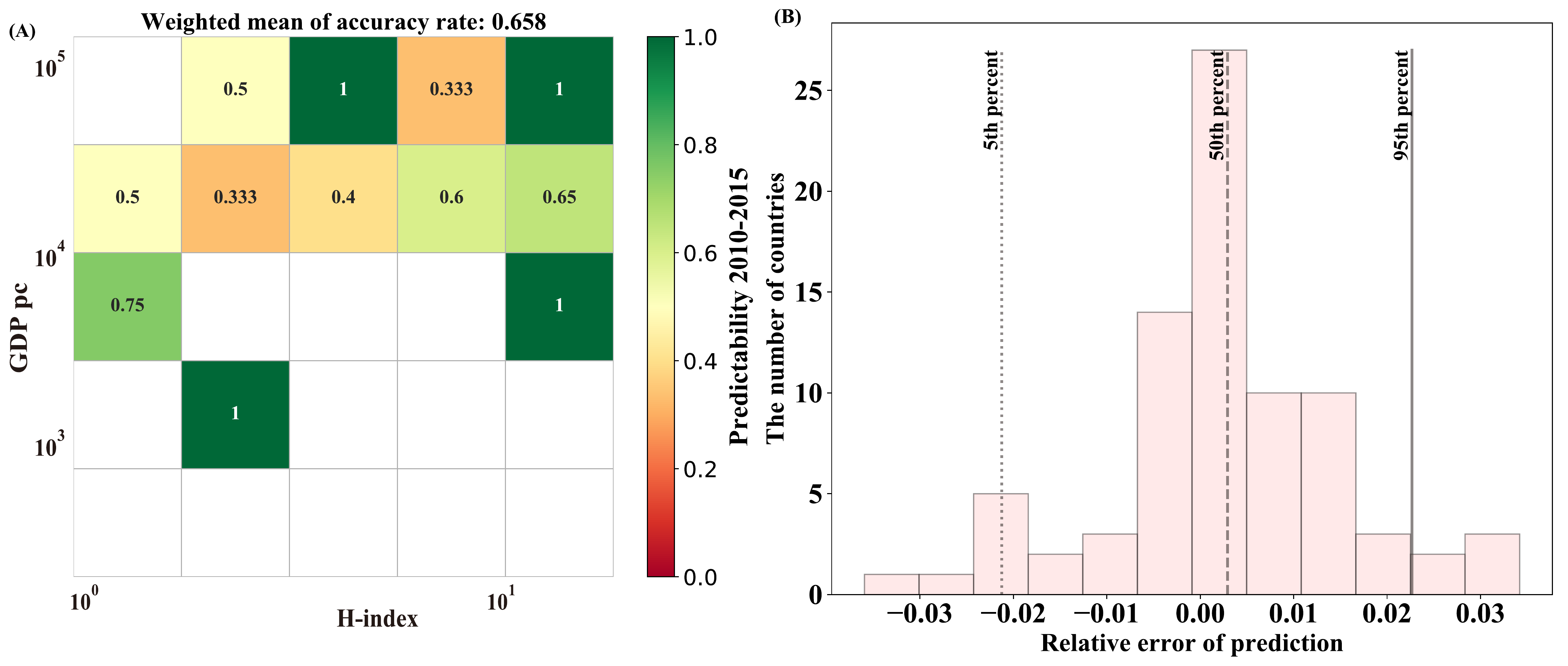} 
\caption{Five-year back-test of Selective Predictability Scheme. 
(A) The accuracy rate of forecasting the located box in the plane of all countries in 2015, the result of each box is measured by the frequency of correct forecasting. The weighted mean of predictive accuracy of the effective boxes is 0.658.
(B) The histogram of the relative errors (see Eq.~(\ref{equation:relative_error})) for all countries in 2015,
which tends to be substantially bell-shaped and peaked around 0. The mean of the relative errors is 0.00216 (S.E. 0.01206).}
\label{figure6}
\end{figure}

To prove the predictability of countries' evolution in the HI-GDP plane, we perform a five-year back-test analysis based on the Selective Predictability Scheme (SPS)~\cite{tacchella2018dynamical} (see Sec.~3.4 for details). Specifically, we use the H-index and GDP pc records of countries from 2000 to 2010 as the training set, which will be split into several 5-year intervals, i.e., $(2000, 2005), (2001, 2006), \ldots, (2005, 2010)$. Then we obtain the moving vector of each country in the HI-GDP plane for each 5-year interval. Further, for each box, we pick up all vectors whose heads are located in this box, and the average displacement of this box is computed as the displacement from the center of chosen vectors' starting positions to the center of all their ending positions.
As a final step, a country's position in the plane in 2015 is estimated as the average displacement of the box it belongs to based on its position in 2010. 
The accuracy results are shown in Fig.~\ref{figure6}. In panel (A) of Fig.~\ref{figure6}, we test the rate of correctly predicting the box in which a country will be located in 2015. The result confirms that countries' dynamics are highly predictive, resulting in 0.658 of the weighted mean of accuracy rate which reduces all the accuracy rate in the plane to a weighted mean value, similar to the calculation of the weighted mean of dispersion.
In Fig.~\ref{figure6} (B), we report the distribution of the relative error between countries' actual and forecasted GDP pc in 2015,
specifically, the relative error of prediction for each country is computed as:
\begin{equation}
Relative~error= \frac{Forecasted~log(GDP~pc)-Actual~log(GDP~pc)}{Actual~log(GDP~pc)}.
\label{equation:relative_error}
\end{equation}
We find that the mean over all countries' relative errors is 0.00216 (S.E. 0.01206), which confirms that the evolution forecast in the HI-GDP plane is accurate for most countries.

We further perform three 5-year predictions, which cover three prediction windows: 2008–2013, 2009–2014 and 2010–2015.
When predicting countries' evolution from $t$ to $t+5$,  
we train the SPS based on countries' dynamics from $t-10$ to $t$.
Subsequently, following the prediction process of SPS described above, we obtain countries' forecasted GDP pc in time $t+5$, 
and compute the compound annual growth rate (CAGR) from $t$ to $t+5$ for each country, which satisfies the equation:  
\begin{equation}
GDP\ pc_{t}*(1+CAGR)^5 = GDP\ pc_{t+5}.
\label{equation: CAGR}
\end{equation}
In this way, each country will have three forecasted CAGR results for three five-year prediction windows, respectively.
We further consider the IMF and fitness forecasting results as baselines, which also cover the corresponding three five-year windows. Therefore, the three considered methods are the new SPS in HI-GDP plane (SPS\_HG), and two baseline methods: IMF and SPS in fitness-GDP plane (SPS\_FG), where IMF predictions are according to various economic data, such as 
inflation rates, current account transactions and foreign trade, which are officially provided by member countries~\cite{dreher2008political}.

To compare the methods' predictive accuracy in terms of CAGR, we calculate the mean absolute error (MAE) and root mean square error (RMSE) of the forecasted CAGR for each country in every window, where the error is defined as~\cite{tacchella2018dynamical}:
$E=|CAGR-CAGR_{forecasted}|$. 
To ensure that there is the same number of predicted instances for the three analyzed methods, we only consider countries for which a prediction by all the three methods is available for all three windows. 
With this filter, we end up with 223 predicted instances.

\begin{table}[ht]
    \centering
    \caption{MAE and RMSE of the GDP pc prediction results. Bold text denotes the best result.}
    \begin{tabular}{ccc}
        \toprule  
        Methods   &   MAE  &  RMSE\\
        \midrule  
        SPS\_HG & \textbf{2.04\%} & \textbf{2.76\%}\\
        IMF & 3.22\% & 3.75\%\\
        SPS\_FG & 2.17\% & 2.80\% \\
        \bottomrule 
    \end{tabular}
    \label{tab:backtest}
\end{table}

The average results over three windows are reported in Table~\ref{tab:backtest}.
We find that SPS in the HI-GDP plane achieves the lowest error rate, it outperforms the IMF predictions by more than 35\% in MAE and 25\% in RMSE, which confirms the predictive value of the countries' technological centrality.
Besides, we confirm Tacchella et al.~\cite{tacchella2018dynamical}'s conclusion that the fitness-based predictions (SPS\_FG) are substantially more accurate than those by the IMF.
At the same time, the predictive error by the SPS\_HG is marginally lower than that by the SPS\_FG for both the two kinds of errors, which suggests that the technological centrality may be a predictor as useful as the economic fitness extracted from the trade data. Besides, the computation of H-index is much simpler compared with the iterative computation process of fitness, exhibiting larger potential in practical applications.

\begin{table}[!ht]
  \centering
  \caption{Top 20 countries with the highest average prediction improvement comparing with the IMF forecasting. CAGR error $E=|CAGR-CAGR_{forecasted}|$. The error results are averaged over three prediction windows. Countries colored in purple are developed countries and the rest are developing countries, according to the IMF's World Economic Outlook Database published in October 2018~\cite{WorldEconomicOutlook2}. See the actual and forecasted CAGR in each window in Table~\ref{tab:avg_error}.}
    \begin{tabular}{lrrr}
     \toprule  
    Country & \multicolumn{1}{l}{Average $E$ of IMF} & \multicolumn{1}{l}{Average $E$ of SPS\_HG} & \multicolumn{1}{l}{SPS\_HG improvement} \\
    \midrule 
    Belarus & 0.055  & 0.009  & 0.046  \\
    Egypt & 0.049  & 0.004  & 0.045  \\
    Tunisia & 0.048  & 0.003  & 0.045  \\
    Ukraine & 0.071  & 0.027  & 0.044  \\
    Russia & 0.050  & 0.009  & 0.041  \\
    Bulgaria & 0.048  & 0.010  & 0.038  \\
    Romania & 0.049  & 0.016  & 0.033  \\
    Samoa & 0.051  & 0.019  & 0.032  \\
    \color{purple}Slovenia  & 0.055  & 0.023  & 0.032  \\
     \color{purple}Slovakia & 0.044  & 0.016  & 0.028  \\
    Senegal & 0.032  & 0.004  & 0.028  \\
    Czech Republic & 0.046  & 0.019  & 0.027  \\
    \color{purple}Sweden  & 0.033  & 0.006  & 0.027  \\
    Hungary & 0.036  & 0.011  & 0.025  \\
    Thailand & 0.028  & 0.004  & 0.024  \\
    Poland & 0.029  & 0.007  & 0.022  \\
    South Africa & 0.041  & 0.019  & 0.022  \\
    Brazil & 0.031  & 0.009  & 0.022  \\
    Mexico & 0.038  & 0.016  & 0.022  \\
    Lebanon & 0.064  & 0.043  & 0.021  \\
     \bottomrule 
    \end{tabular}
  \label{tab:enhancements}
\end{table}

\begin{figure}[!ht]
\centering
\includegraphics[scale=0.5]{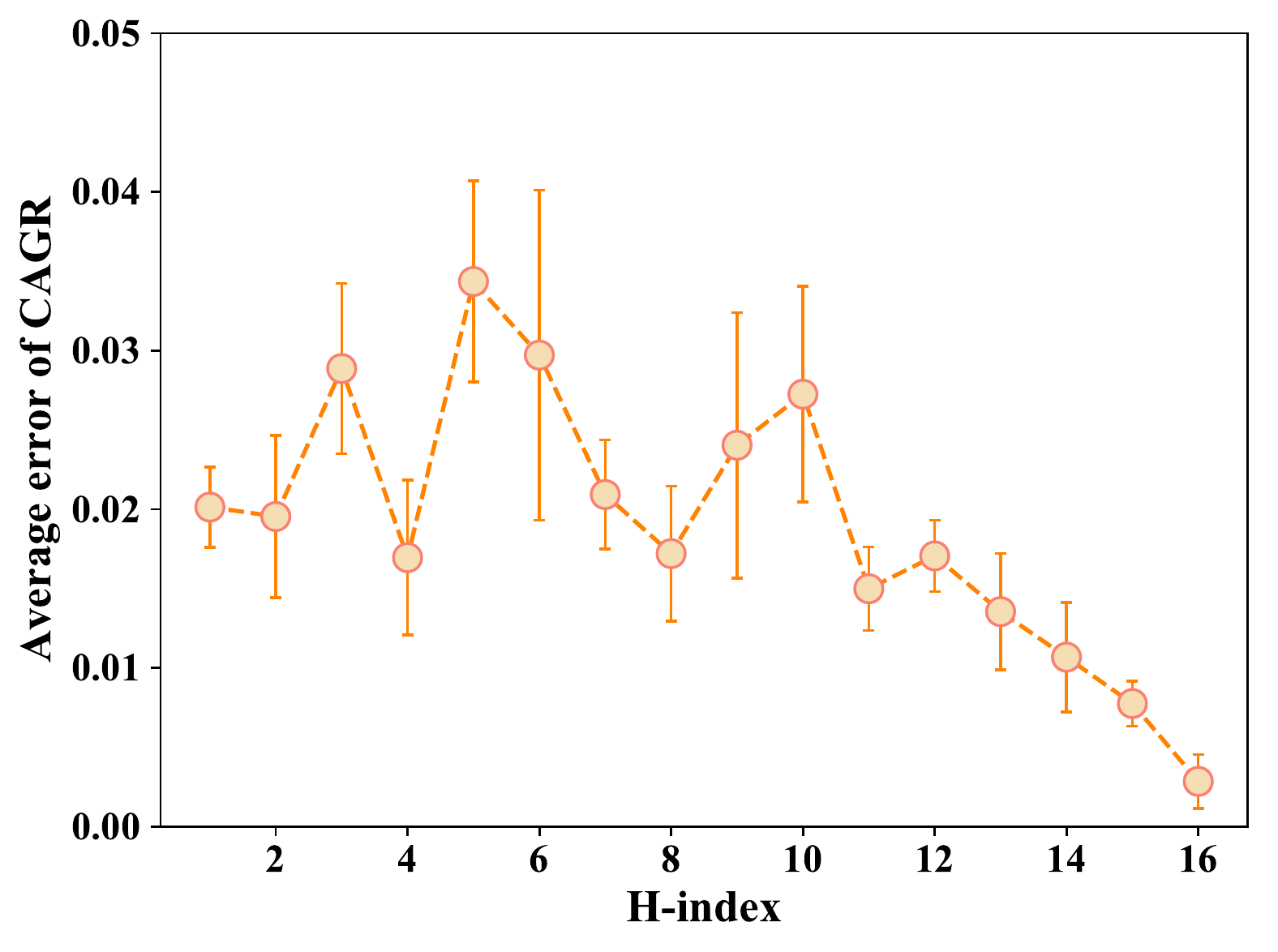} 
\caption{Average prediction error of countries with different H-index.
We show the average prediction error of countries with different H-index for the three five-year forecasts. Dots represent the average error and error bars represent the standard error of the mean.}
\label{figure7}
\end{figure}

In Table~\ref{tab:enhancements}, we list the top 20 countries by the highest average prediction improvement comparing with the IMF forecasting. The prediction improvement is defined as the difference of average results between the CAGR errors $E$ of SPS in the HI-GDP plane and those of IMF over the three prediction windows.
As can be seen from the table, 17 of the top 20 countries are developing countries, which suggests that the SPS method based on the HI-GDP plane is more accurate in forecasting economic development for developing countries.
To further verify the stability of the method's predictive performance, we examine the dependency of the country-level prediction error in the HI-GDP plane in Fig.~\ref{figure7}.
We find that countries with high H-index have smaller average CAGR error, which suggests that the more technologically advanced a country is, the more predictable its future GDP pc is. 

\section{Discussion and conclusion}

Departing from standard approaches in economics, the recent economic complexity literature has revealed that low-dimensional predictive models for a country's gross domestic product can perform similarly well or even better than high-dimensional ones, such as the one by the IMF~\cite{tacchella2018dynamical}. At the same time, virtually all studies in the field have focused on inferring a country's economic fitness from export data. Departing from this focus, our work proposed a different approach to forecast the GDP pc based on the technological centrality of countries, inferred from patent data.
By constructing a patent citation network between countries, we measured various network centralities and demonstrated a significant positive correlation between the centrality metrics (especially the H-index) and the GDP pc. 
We further revealed the strong predictability of the GDP pc based on the countries' technological H-index by tracing the countries' evolution in the HI-GDP plane.
We built a prediction model based on the HI-GDP plane, and found that the resulting predictions substantially outperform those by the IMF and marginally outperform those by the economic fitness method.

There are exciting research opportunities opened by our study and its limitations.
Firstly, a potential limitation of our study is that it only includes data from the European Patent Office, thus it may lead to the underestimation of the applicants who resides in non-European countries due to the \textit{home bias effect}. Overcoming this barrier is an important direction for future research, and it might involve merging data from multiple patent offices. Secondly, the correlation between countries' technological centrality and future economic development, together with the resulting predictive power, does not imply that there is a causation between the two variables. More work is needed to understand and clarify the underlying mechanism behind the observed correlation and predictive power. Thirdly, more research is needed to understand the reason behind H-index's advantage over other centrality metrics and to compare the predictive ability of the technological centrality metrics with centrality metrics extracted from different kinds of technological and scientific networks. Finally, the proposed method is not limited to the study of nations, it can be extended to the investigation of regions, cities as well.

To conclude, despite its potential limitations, our study provides a first attempt toward predicting the GDP of nations based on a single indicator that is not extracted from export data. The observed predictability not only can help to develop effective economic-forecasting indicators and inspire more research in economic complexity, but it can also deepen our understanding of economic development and deliver valuable insights to governments. 

\section*{Acknowledgements}
We thank Luciano Pietronero, Emanuele Pugliese, and Andrea Zaccaria for providing us with the patent data and enlightening discussions.
The work was supported by the National Natural Science Foundation of China (Grant Nos.11622538), the XPLORER PRIZE, and the Special Project for the Central Guidance on Local Science and Technology Development of Sichuan Province (2021ZYD0029). M.S.M. acknowledges financial support from the URPP Social Networks at the University of Zurich, and the UESTC professor research start-up (Grant No. ZYGX2018KYQD215).

\bibliography{ChaosSF_draft}
\section*{Appendix}
\appendix 
\setcounter{table}{0} 
\setcounter{figure}{0}

\renewcommand{\thetable}{A\arabic{table}}
\renewcommand{\thefigure}{A\arabic{figure}}

\begin{table}[!ht]
    \centering
    \footnotesize
    \caption{The ISO 3166 international standard code for countries/regions involved in the analysis.}
    \begin{tabular}{p{1.2cm}p{2.8cm}p{1.2cm}p{2.8cm}p{1.2cm}p{2.8cm}}
    \toprule
    \multicolumn{2}{c|}{European} & \multicolumn{4}{c}{Non-European} \\
     \midrule
    Code  & \multicolumn{1}{l|}{Country or region} & Code  & \multicolumn{1}{l}{Country or region} & Code  & \multicolumn{1}{l}{Country or region} \\
    \midrule
    AD    & \multicolumn{1}{l|}{Andorra} & AE    & Arab Emirates & LR    & Liberia \\
    AT    & \multicolumn{1}{l|}{Austria} & AG    & Antigua and Barb. & LY    & Libya \\
    BE    & \multicolumn{1}{l|}{Belgium} & AN    & Antilles & MA    & Morocco \\
    BG    & \multicolumn{1}{l|}{Bulgaria} & AR    & Argentina & MH    & Marshall Islands \\
    BM    & \multicolumn{1}{l|}{Bermuda} & AU    & Australia & ML    & Mali \\
    BY    & \multicolumn{1}{l|}{Belarus} & AZ    & Azerbaijan & MO    & Macau, China \\
    CH    & \multicolumn{1}{l|}{Switzerland} & BB    & Barbados & MU    & Mauritius \\
    CZ    & \multicolumn{1}{l|}{Czech Republic} & BH    & Bahrain & MX    & Mexico \\
    DE    & \multicolumn{1}{l|}{Germany} & BN    & Brunei & MY    & Malaysia \\
    DK    & \multicolumn{1}{l|}{Denmark} & BR    & Brazil & NZ    & New Zealand \\
    EE    & \multicolumn{1}{l|}{Estonia} & BS    & Bahamas & PA    & Panama \\
    ES    & \multicolumn{1}{l|}{Spain} & BZ    & Belize & PE    & Peru \\
    FI    & \multicolumn{1}{l|}{Finland} & CA    & Canada & PH    & Philippines \\
    FR    & \multicolumn{1}{l|}{France} & CK    & Cook Islands & PK    & Pakistan \\
    GB    & \multicolumn{1}{l|}{United Kingdom} & CL    & Chile & PR    & Puerto Rico \\
    GI    & \multicolumn{1}{l|}{Gibraltar} & CN    & China & QA    & Qatar \\
    GR    & \multicolumn{1}{l|}{Greece} & CO    & Colombia & RH    & Rhodesia \\
    HR    & \multicolumn{1}{l|}{Croatia} & CR    & Costa Rica & SA    & Saudi Arabia \\
    HU    & \multicolumn{1}{l|}{Hungary} & CU    & Cuba  & SC    & Seychelles \\
    IE    & \multicolumn{1}{l|}{Ireland} & CW    & Curaçao & SG    & Singapore \\
    IM    & \multicolumn{1}{l|}{Isle of Man} & CY    & Cyprus & SL    & Sierra Leone \\
    IS    & \multicolumn{1}{l|}{Iceland} & DM    & Dominica & SN    & Senegal \\
    IT    & \multicolumn{1}{l|}{Italy} & DO    & Dominican Republic & SY & Syria \\
    LI    & \multicolumn{1}{l|}{Liechtenstein} & DZ    & Algeria & TC & Turks and Caicos Islands \\
    LT    & \multicolumn{1}{l|}{Lithuania} & EC    & Ecuador & TH & Thailand \\
    LU    & \multicolumn{1}{l|}{Luxembourg} & EG    & Egypt & TN & Tunisia \\
    LV    & \multicolumn{1}{l|}{Latvia} & GA    & Gabon & \multicolumn{1}{l}{TR} & \multicolumn{1}{l}{Turkey} \\
      MC    & \multicolumn{1}{l|}{Monaco} & GE    & Georgia & \multicolumn{1}{l}{TW} & \multicolumn{1}{l}{Taiwan, China} \\
         ME    & \multicolumn{1}{l|}{Montenegro} & GT    & Guatemala & \multicolumn{1}{l}{US} & \multicolumn{1}{l}{United States} \\
       MT    & \multicolumn{1}{l|}{Malta} & HK    & Hong Kong, China & \multicolumn{1}{l}{UY} & \multicolumn{1}{l}{Uruguay} \\
         NL    & \multicolumn{1}{l|}{Netherlands} & ID    & Indonesia & \multicolumn{1}{l}{UZ} & \multicolumn{1}{l}{Uzbekistan} \\
        NO    & \multicolumn{1}{l|}{Norway} & IL    & Israel & \multicolumn{1}{l}{VC} & \multicolumn{1}{l}{Saint Vincent and the Grenadines} \\
     PL    & \multicolumn{1}{l|}{Poland} & IN    & India & \multicolumn{1}{l}{VE} & \multicolumn{1}{l}{Venezuela} \\
       PT    & \multicolumn{1}{l|}{Portugal} & IR    & Iran  & \multicolumn{1}{l}{VG} & \multicolumn{1}{l}{British Virgin Islands} \\
    RE    & \multicolumn{1}{l|}{Reunion} & JO    & Jordan & \multicolumn{1}{l}{VI} & \multicolumn{1}{l}{United States Virgin Islands} \\
     RO    & \multicolumn{1}{l|}{Romania} & JP    & Japan & \multicolumn{1}{l}{VN} & \multicolumn{1}{l}{Vietnam} \\
    RS    & \multicolumn{1}{l|}{Serbia} & KP    & Democratic People's Republic of Korea & \multicolumn{1}{l}{WS} & \multicolumn{1}{l}{Samoa} \\
     RU    & \multicolumn{1}{l|}{Russia} & KR    & Republic of Korea & \multicolumn{1}{l}{ZA} & \multicolumn{1}{l}{South Africa} \\
    SE    & \multicolumn{1}{l|}{Sweden} & KW    & Kuwait & \multicolumn{1}{l}{ZW} & \multicolumn{1}{l}{Zimbabwe} \\
    SI    & \multicolumn{1}{l|}{Slovenia} & KY    & Cayman Islands &       &  \\
    SK    & \multicolumn{1}{l|}{Slovakia} & KZ    & Kazakhstan &       &  \\
    SM    & \multicolumn{1}{l|}{San Marino} & LB    & Lebanon &       &  \\
    UA    & \multicolumn{1}{l|}{Ukraine} & LK    & Sri Lanka &       &  \\
    \midrule
   \multicolumn{6}{c}{Special Case} \\
    \midrule
    Code  & \multicolumn{1}{l}{Country or region} & \multicolumn{1}{l|}{Remark}  & Code & Organization & Remark \\
      \midrule
      CS    & Czechoslovakia & \multicolumn{1}{l|}{Dissolved } & \multicolumn{1}{l}{EP} & \multicolumn{1}{l}{EPO} & \multicolumn{1}{l}{Patent organization} \\
    DD    & German Democratic Republic & \multicolumn{1}{l|}{Dissolved } & \multicolumn{1}{l}{WO} & WIPO & Patent organization \\
    YU    & Yugoslavia & \multicolumn{1}{l|}{Dissolved }  &       &       &  \\
    SU    & Soviet Union &    \multicolumn{1}{l|}{Dissolved }   &       &       &  \\
    \bottomrule
    \end{tabular}
\end{table}

\begin{table}[!h]
  \centering
  \caption{The normalized H-index of countries/regions in each year from 1995 to 2015. Countries or regions in the table are from the union of the Top 20 performers ranked by H-index in each year.}
  \footnotesize
    \begin{tabular}{p{1.5cm}p{0.25cm}p{0.25cm}p{0.25cm}p{0.25cm}p{0.25cm}p{0.25cm}p{0.25cm}p{0.25cm}p{0.25cm}p{0.25cm}p{0.25cm}p{0.25cm}p{0.25cm}p{0.25cm}p{0.25cm}p{0.25cm}p{0.25cm}p{0.25cm}p{0.25cm}p{0.25cm}p{0.25cm}}
    \toprule
    Year  & 1995  & 1996  & 1997  & 1998  & 1999  & 2000  & 2001  & 2002  & 2003  & 2004  & 2005  & 2006  & 2007  & 2008  & 2009  & 2010  & 2011  & 2012  & 2013  & 2014  & 2015 \\
    \midrule
    Germany & 2.72  & 2.69  & 2.69  & 2.72  & 2.70   & 2.84  & 2.80   & 2.88  & 2.83  & 2.82  & 2.77  & 2.86  & 2.76  & 2.67  & 2.65  & 2.73  & 2.66  & 2.58  & 2.54  & 2.45  & 2.47 \\
    Luxembourg & 2.51  & 2.27  & 2.07  & 2.09  & 2.29  & 2.63  & 2.40   & 2.47  & 2.42  & 2.82  & 2.77  & 2.67  & 2.58  & 2.31  & 2.12  & 2.22  & 2.49  & 2.41  & 2.38  & 2.45  & 2.47 \\
    United States & 2.30   & 2.48  & 2.49  & 2.51  & 2.50   & 2.63  & 2.60   & 2.67  & 2.63  & 2.62  & 2.57  & 2.28  & 2.21  & 2.31  & 2.30   & 2.22  & 2.32  & 2.41  & 2.38  & 2.14  & 2.32 \\
    Sweden & 2.51  & 2.48  & 2.49  & 2.30   & 2.50   & 2.63  & 2.60   & 2.67  & 2.63  & 2.62  & 2.57  & 2.48  & 2.39  & 2.49  & 2.48  & 2.73  & 2.66  & 2.58  & 2.54  & 2.30   & 2.32 \\
    Portugal & 0.21  & 0.62  & 0.41  & 0.63  & 0.62  & 0.81  & 1.20   & 1.23  & 1.41  & 1.61  & 1.78  & 1.90   & 1.84  & 1.78  & 1.95  & 1.88  & 1.66  & 1.61  & 1.75  & 1.99  & 2.32 \\
    France & 2.51  & 2.69  & 2.69  & 2.72  & 2.50   & 2.63  & 2.60   & 2.67  & 2.22  & 2.22  & 2.18  & 2.28  & 2.21  & 2.13  & 2.30   & 2.39  & 2.16  & 2.25  & 2.06  & 2.14  & 2.16 \\
    Canada & 2.30   & 2.27  & 2.28  & 2.30   & 2.08  & 2.03  & 2.00     & 2.06  & 1.82  & 2.02  & 1.98  & 1.90   & 2.03  & 2.13  & 2.12  & 2.22  & 2.16  & 2.09  & 2.06  & 1.99  & 2.16 \\
     Denmark & 2.51  & 2.27  & 2.49  & 2.30   & 2.29  & 2.43  & 2.40   & 2.67  & 2.22  & 2.42  & 2.37  & 2.48  & 2.39  & 2.49  & 2.48  & 2.56  & 2.16  & 1.93  & 2.06  & 1.99  & 2.16 \\
    Japan & 1.67  & 1.65  & 1.66  & 1.88  & 1.66  & 1.62  & 1.60   & 1.65  & 1.82  & 2.02  & 1.98  & 2.09  & 2.21  & 2.13  & 2.30   & 2.39  & 2.16  & 1.93  & 1.91  & 1.99  & 2.16 \\
    Spain & 2.09  & 2.07  & 1.86  & 1.88  & 2.29  & 2.03  & 2.20   & 2.26  & 2.22  & 2.22  & 2.37  & 2.48  & 2.21  & 2.13  & 2.30   & 2.05  & 2.16  & 2.25  & 2.06  & 2.14  & 2.16 \\
    Belgium & 2.09  & 2.07  & 2.07  & 2.09  & 2.29  & 2.23  & 2.00     & 1.85  & 2.02  & 2.02  & 1.98  & 1.90   & 1.84  & 2.13  & 1.95  & 1.88  & 1.83  & 1.93  & 1.91  & 1.84  & 2.16 \\
    Australia & 2.09  & 2.07  & 2.07  & 2.09  & 2.08  & 2.23  & 2.20   & 2.26  & 2.22  & 2.22  & 2.37  & 2.28  & 2.21  & 2.31  & 2.12  & 2.22  & 2.16  & 2.25  & 2.22  & 2.30   & 2.16 \\
    Israel & 2.09  & 2.07  & 1.86  & 2.09  & 1.87  & 1.82  & 2.00     & 1.65  & 1.82  & 2.02  & 1.98  & 1.90   & 1.84  & 1.96  & 1.77  & 2.05  & 1.99  & 1.93  & 1.91  & 1.99  & 2.16 \\
    Norway & 2.09  & 2.07  & 2.07  & 2.09  & 1.87  & 1.82  & 2.00     & 1.85  & 1.82  & 1.81  & 1.98  & 1.90   & 2.21  & 2.13  & 2.12  & 2.05  & 2.16  & 2.09  & 2.22  & 1.99  & 2.16 \\
     Hong Kong, China & 0.84  & 1.03  & 0.83  & 0.84  & 1.25  & 1.22  & 1.20   & 1.23  & 1.41  & 1.41  & 1.58  & 1.71  & 2.03  & 2.13  & 1.95  & 2.05  & 2.16  & 1.93  & 2.06  & 1.99  & 2.16 \\
         Ireland & 2.30   & 2.07  & 2.28  & 2.09  & 1.87  & 2.23  & 2.20   & 2.26  & 2.22  & 2.22  & 2.37  & 2.28  & 2.21  & 2.31  & 2.12  & 2.22  & 2.16  & 2.41  & 2.22  & 2.14  & 2.16 \\
    Italy & 1.88  & 2.07  & 1.66  & 1.88  & 1.87  & 2.03  & 2.00     & 2.06  & 2.02  & 2.02  & 1.98  & 2.09  & 2.03  & 2.13  & 2.12  & 2.05  & 1.99  & 2.09  & 2.06  & 1.99  & 2.16 \\
    China & 1.05  & 1.24  & 1.04  & 1.05  & 1.25  & 1.42  & 1.40   & 1.44  & 1.41  & 1.61  & 1.58  & 1.71  & 1.84  & 2.13  & 1.95  & 1.88  & 1.99  & 1.93  & 2.06  & 1.99  & 2.01 \\
    Poland & 0.84  & 0.83  & 0.62  & 0.63  & 0.62  & 0.61  & 0.60   & 0.82  & 0.81  & 0.81  & 0.59  & 0.57  & 0.92  & 0.89  & 0.88  & 1.19  & 1.16  & 1.29  & 1.59  & 1.84  & 2.01 \\
    Greece & 0.84  & 0.62  & 1.04  & 1.26  & 1.25  & 1.01  & 1.20   & 1.23  & 1.21  & 1.41  & 1.58  & 1.33  & 1.47  & 1.60   & 1.77  & 1.70   & 1.83  & 2.09  & 1.91  & 1.84  & 2.01 \\
    Switzerland & 1.88  & 2.07  & 2.07  & 1.88  & 1.87  & 2.03  & 2.00     & 2.06  & 2.02  & 2.02  & 1.98  & 1.90   & 2.03  & 2.13  & 2.12  & 2.05  & 1.99  & 1.93  & 2.06  & 1.99  & 2.01 \\
    Austria & 2.30   & 2.07  & 2.28  & 2.09  & 1.87  & 2.03  & 2.00     & 2.06  & 2.02  & 2.02  & 1.78  & 1.90   & 2.03  & 2.13  & 1.95  & 1.88  & 1.83  & 1.77  & 1.91  & 1.99  & 2.01 \\
    Hungary & 1.88  & 1.86  & 1.86  & 1.88  & 1.87  & 2.03  & 2.00     & 2.26  & 2.22  & 2.22  & 2.18  & 2.09  & 2.21  & 2.13  & 2.12  & 2.22  & 1.99  & 2.09  & 1.91  & 1.99  & 1.85 \\
    Netherlands & 1.88  & 1.86  & 1.86  & 2.09  & 1.87  & 1.82  & 2.00     & 1.65  & 1.82  & 1.81  & 1.98  & 1.90   & 1.84  & 1.78  & 1.77  & 1.88  & 1.83  & 1.93  & 1.91  & 1.84  & 1.85 \\
    United Kingdom & 2.09  & 2.07  & 2.07  & 2.3   & 2.08  & 2.23  & 2.00     & 2.06  & 2.02  & 2.02  & 1.78  & 1.90   & 1.84  & 1.96  & 1.95  & 1.88  & 1.99  & 1.93  & 1.91  & 1.84  & 1.85 \\
    Panama & 1.47  & 1.45  & 1.66  & 1.67  & 1.87  & 1.62  & 1.80   & 1.65  & 1.82  & 1.81  & 1.98  & 1.90   & 1.84  & 2.13  & 1.95  & 2.05  & 1.99  & 1.93  & 1.91  & 1.84  & 1.85 \\
    Slovenia & 1.05  & 1.03  & 1.04  & 0.84  & 1.04  & 1.22  & 1.20   & 1.44  & 1.62  & 1.61  & 1.78  & 1.90   & 2.03  & 1.96  & 1.95  & 1.88  & 1.83  & 1.77  & 1.91  & 1.99  & 1.85 \\
     Finland & 1.88  & 1.86  & 1.86  & 1.88  & 1.66  & 1.62  & 2.00     & 1.65  & 1.62  & 1.81  & 1.98  & 2.09  & 2.21  & 1.96  & 2.12  & 1.88  & 1.83  & 1.77  & 1.75  & 1.84  & 1.85 \\
    New Zealand & 1.47  & 1.45  & 1.45  & 1.26  & 1.46  & 1.42  & 1.60   & 1.23  & 1.41  & 1.61  & 1.78  & 1.71  & 1.66  & 1.96  & 1.59  & 1.88  & 1.99  & 1.77  & 1.75  & 1.84  & 1.70 \\
    Korea & 1.05  & 0.62  & 0.41  & 0.63  & 1.04  & 1.01  & 1.20   & 1.03  & 1.21  & 1.41  & 1.78  & 1.90   & 2.03  & 2.13  & 2.12  & 1.88  & 1.99  & 1.77  & 1.75  & 1.68  & 1.54 \\
    \bottomrule
    \end{tabular}
  \label{tab:norm_H}
\end{table}

\begin{sidewaystable}[!h]
\footnotesize
	\centering
		\caption{Top 20 countries with the highest average prediction improvement comparing with the IMF forecasting. CAGR error $E=|CAGR-CAGR_{forecasted}|$. The error results are averaged over three prediction windows.}
    \begin{tabular}{lrrrrrrrrrrrr}
    \toprule
          & \multicolumn{3}{c}{CAGR} & \multicolumn{3}{c}{CAGR$_{forecasted}$ by IMF } & \multicolumn{3}{c}{CAGR$_{forecasted}$ by SPS\_HG } & \multicolumn{1}{l}{Avg $E$ of IMF} & \multicolumn{1}{l}{Avg $E$ of SPS\_HG} & \multicolumn{1}{l}{SPS\_HG improvement} \\
    Country & \multicolumn{1}{l}{ 2008-2013} & \multicolumn{1}{l}{ 2009-2014} & \multicolumn{1}{l}{2010-2015} & \multicolumn{1}{l}{2008-2013} & \multicolumn{1}{l}{2009-2014} & \multicolumn{1}{l}{2010-2015} & \multicolumn{1}{l}{2008-2013} & \multicolumn{1}{l}{2009-2014} & \multicolumn{1}{l}{2010-2015} &       &       &  \\
    \midrule
    Belarus & 0.033  & 0.036  & 0.012  & 0.092  & 0.081  & 0.071  & 0.032  & 0.034  & 0.035  & 0.055  & 0.009  & 0.046  \\
    Egypt & 0.011  & 0.006  & 0.004  & 0.062  & 0.051  & 0.055  & 0.009  & 0.009  & 0.012  & 0.049  & 0.004  & 0.045  \\
    Tunisia & 0.013  & 0.013  & 0.008  & 0.059  & 0.060  & 0.059  & 0.009  & 0.009  & 0.012  & 0.048  & 0.003  & 0.045  \\
    Ukraine & -0.010  & 0.019  & -0.009  & 0.076  & 0.072  & 0.066  & 0.025  & 0.027  & 0.029  & 0.071  & 0.027  & 0.044  \\
    Russia & 0.011  & 0.026  & 0.012  & 0.081  & 0.057  & 0.062  & 0.027  & 0.025  & 0.024  & 0.050  & 0.009  & 0.041  \\
    Bulgaria & 0.006  & 0.017  & 0.024  & 0.082  & 0.050  & 0.061  & 0.027  & 0.025  & 0.024  & 0.048  & 0.010  & 0.038  \\
    Romania & 0.001  & 0.018  & 0.034  & 0.077  & 0.069  & 0.055  & 0.032  & 0.034  & 0.035  & 0.049  & 0.016  & 0.033  \\
    Samoa & -0.018  & -0.008  & 0.000  & 0.054  & 0.033  & 0.040  & 0.009  & 0.009  & 0.012  & 0.051  & 0.019  & 0.032  \\
    Slovenia & -0.022  & 0.000  & 0.003  & 0.057  & 0.046  & 0.043  & 0.019  & 0.017  & 0.015  & 0.055  & 0.023  & 0.032  \\
    Slovakia & 0.009  & 0.026  & 0.025  & 0.073  & 0.061  & 0.058  & 0.037  & 0.037  & 0.034  & 0.044  & 0.016  & 0.028  \\
    Senegal & 0.002  & 0.011  & 0.016  & 0.049  & 0.038  & 0.038  & 0.009  & 0.009  & 0.012  & 0.032  & 0.004  & 0.028  \\
    Czech Republic & -0.007  & 0.009  & 0.015  & 0.060  & 0.048  & 0.048  & 0.027  & 0.025  & 0.024  & 0.046  & 0.019  & 0.027  \\
    Sweden & 0.002  & 0.016  & 0.013  & 0.043  & 0.041  & 0.044  & 0.020  & 0.016  & 0.012  & 0.033  & 0.006  & 0.027  \\
    Hungary & -0.005  & 0.017  & 0.024  & 0.051  & 0.048  & 0.045  & 0.019  & 0.017  & 0.015  & 0.036  & 0.011  & 0.025  \\
    Thailand & 0.030  & 0.033  & 0.025  & 0.063  & 0.058  & 0.050  & 0.032  & 0.034  & 0.035  & 0.028  & 0.004  & 0.024  \\
    Poland & 0.029  & 0.031  & 0.031  & 0.066  & 0.054  & 0.057  & 0.037  & 0.037  & 0.024  & 0.029  & 0.007  & 0.022  \\
    South Africa & 0.004  & 0.010  & 0.006  & 0.053  & 0.044  & 0.044  & 0.027  & 0.025  & 0.024  & 0.041  & 0.019  & 0.022  \\
    Brazil & 0.023  & 0.024  & 0.003  & 0.045  & 0.045  & 0.050  & 0.027  & 0.025  & 0.024  & 0.031  & 0.009  & 0.022  \\
    Mexico & 0.002  & 0.019  & 0.016  & 0.046  & 0.055  & 0.049  & 0.037  & 0.025  & 0.024  & 0.038  & 0.016  & 0.022  \\
    Lebanon & 0.006  & -0.018  & -0.035  & 0.053  & 0.047  & 0.045  & 0.025  & 0.027  & 0.029  & 0.064  & 0.043  & 0.021  \\
     \bottomrule
     \end{tabular}
     \label{tab:avg_error}
\end{sidewaystable}

\clearpage

\end{document}